\documentclass[11pt,a4paper]{article}

\usepackage{graphicx, amsmath, pifont, comment, layout, amsthm, amssymb}
\usepackage{listings}
\usepackage{eso-pic}
\usepackage{booktabs}
\usepackage{float}

\usepackage[english]{babel}
\usepackage{hyperref} 
\hypersetup{ colorlinks=true, linkcolor=black, filecolor=black, urlcolor=cyan, }

\usepackage{caption}
\makeatletter
\newcommand\notsotiny{\@setfontsize\notsotiny\@vipt\@viipt}
\makeatother

\begin{document}

\title{Lightweight Selective Disclosure for Verifiable Documents on Blockchain}
\author{
Kenji Saito\footnote{Graduate School of Business and Finance, Waseda University},
$\;$
Satoki Watanabe\footnote{Graduate School of Media and Governance, Keio University}
}
\date{}

\maketitle

\begin{abstract}
To achieve lightweight selective disclosure for protecting privacy of
document holders, we propose an XML format for documents that can hide
arbitrary elements using a cryptographic hash function and salts, which allows
to be partially digitally signed and efficiently verified, as well as a JSON
format that can be converted to such XML.
The documents can be efficiently proven to exist by representing multiple such
structures as a Merkle tree and storing its root in blockchain.

We show that our proposal has advantages over known methods that represent
the document itself as a Merkle tree and partially hide it.

\begin{description}
\item[Keywords:]
Selective disclosure, Content extraction signature,\\
Privacy, Proof of existence, Blockchain
\end{description}

\end{abstract}

\section{Introduction}\label{sec-intro}
\subsection{Overview of the Problem}
Selective disclosure is an important mechanism to protect the privacy of users
by hiding unnecessary parts of certificates or other documents, and providing
just partial disclosure for proof.

For example, when buying alcoholic beverages at a shop, you need to prove
that you have reached a certain age.
But many certificates, such as driver's licenses, issued by public institutions
contain more information than just your date of birth.
From the perspective of privacy protection, you should be able to provide
the shop with only the date of birth (or that you are over a certain age),
and to make it verifiable that the fact is certified by a public authority
({\em partial disclosure}).

As another example, when presenting university transcripts to companies for
employment, it should be unnecessary to disclose grades for courses that are
not relevant.
In addition, it is better to have the digital signature of the lecturer for
each course grade to ensure authenticity of evaluation
({\em partial signature}).

\subsection{Contributions}
In this paper, we show a simple solution for these problems.
While the known practice described in
section \ref{sec-known-practice}
allows partial disclosure only,
our proposed technique allows both partial disclosure and partial signature.
We also show that our solution works with less computational overhead with
respect to the number of digest calculations.

\subsection{Limitation}
The correlation problem of possible leakage of identities remains with
the proposed technique,
because our solution has linkability, and the verifier would know
if multiple verifications deal with the same document.
However, this problem can be solved by following a technological standard or
by some legal order.
We will discuss this topic later in section \ref{sec-overcome}.

\section{Background}\label{sec-background}

\subsection{Selective Disclosure}
Selective disclosure has been studied for years, and is sometimes called
{\em content extraction signature} because it extracts some signed verifiable
content.

Other than the simplest solution of signing for each unit that can be
disclosed, there are two approaches for solutions:
a) signature methods that allow for partial
disclosure (e.g. \cite{10.1007/978-3-540-28628-8_4}), and
b) applying cryptographic hash functions to non-disclosed portions
(e.g. \cite{10.1145/775152.775176}).
Our solution is an example of the latter.

Many studies include {\em unlinkability} in their goals, which is to avoid use
of information gathered from each disclosure reaching the identity of the
document holder, making them difficult challenges.

\subsection{Merkle Tree}
Merkle tree\cite{Merkle1988:Tree} is a hash tree structure that allows 
representation of multiple elements with a single value, which is used for
proof of existence of elements while obscuring others.
Our use of Merkle tree is illustrated in Figure \ref{fig-merkle}.

\subsection{Blockchain}
Blockchain\cite{Nakamoto2008:Bitcoin} is a structure with
internal use of Merkle trees and other
cryptographic digest techniques that allows for tamper-evident storage
verifiable by the public.
Verifiability of information in
blockchain is based on replication of state machines to participating nodes.
Analyses of this include \cite{ZHANG202093} and \cite{Saito2016:Blockchain}.

Ethereum\cite{Buterin2013:Ethereum} is a blockchain-based application platform
to assure authenticity of program codes ({\em smart contracts}), their
execution logs and the resulted states.

\subsection{Known Selective Disclosure Practice}\label{sec-known-practice}
There is a known practice of selective disclosure for documents using
blockchain, described, for example, in a blog entry\cite{Hitchens2018:Merkle}.
In the method, a document itself is represented as a Merkle tree by placing
all elements of the document as leaves, and its root is stored in blockchain.

\section{Design}\label{sec-design}

\subsection{XML Data Format}
\begin{figure*}[h!]
\begin{center}
\includegraphics[scale=0.6]{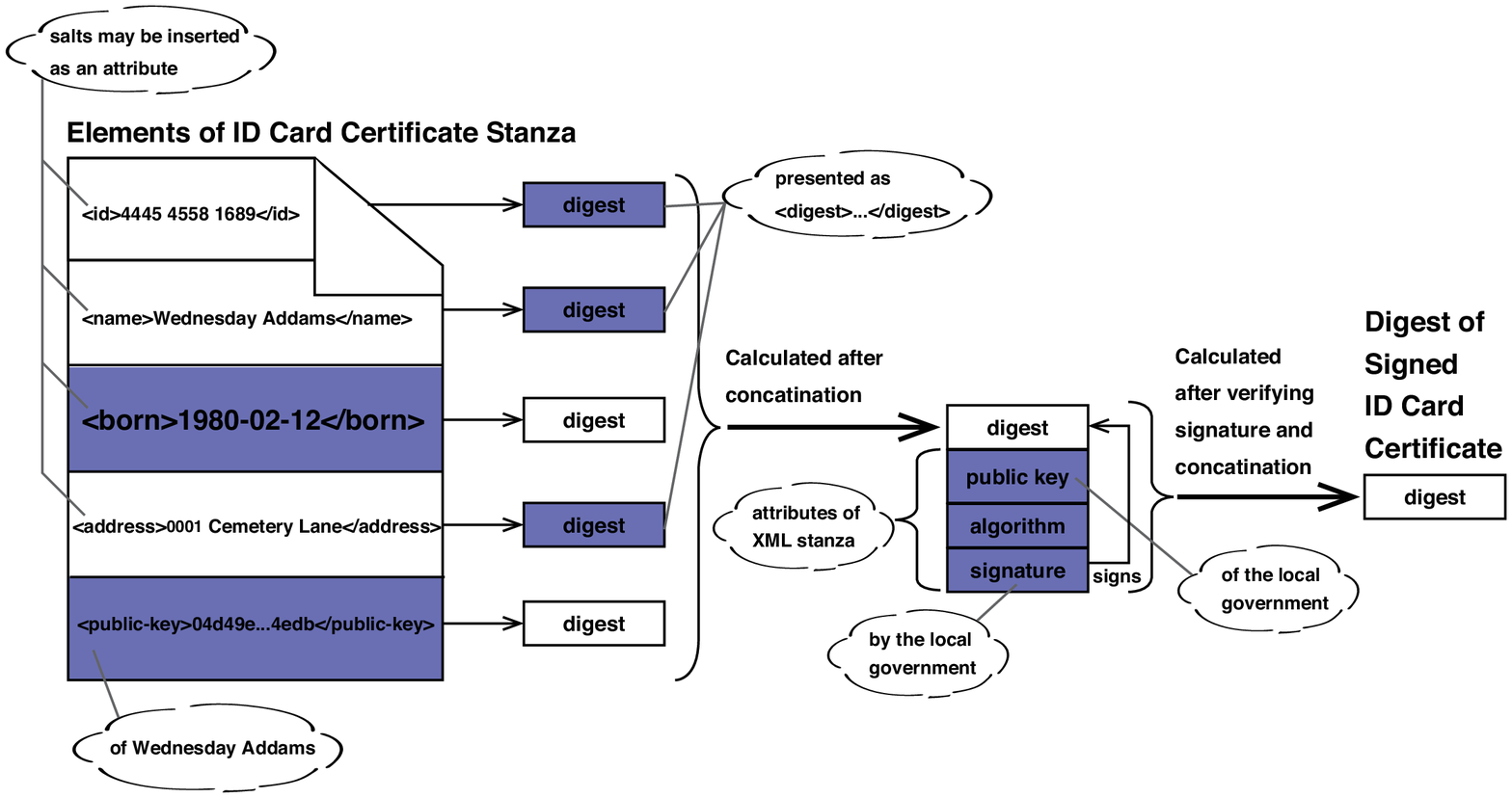}\\
{\footnotesize
\begin{itemize}
\item[*] By providing only the colored parts to the verifier, the structure
can be reproduced and verified, without revealing her ID, name or address.
\item[*] This can be applied recursively to 
any digest calculations (``$\rightarrow$ \framebox{digest} '' on the left) to
make a nested structure that
can be partially signed by relevant parties.
\end{itemize}
}
\caption{Signed XML data format for selective disclosure and how it is
verified.}\label{fig-xml}
\end{center}
\end{figure*}

We propose to express a (signed part of) document as a sequence of XML elements
contained in a container XML stanza, as illustrated in Figure \ref{fig-xml}
(this example is an ID card of Wednesday Addams).
The
container
stanza can have {\em algo}, {\em sig} and {\em pubkey} attributes that
represent the digital signature algorithm, signature and public key for
verifying the signature, respectively.
Each element can have {\em salt} attribute to make it difficult to guess the
preimage of the digest.

We define the cryptographic hash function $h$ for obtaining the digest of a
signed document as follows.
First, we define the basic cryptographic hash function $h'$ as, for example,
SHA-256.
To get the digest for each element in the container stanza, we apply $h$ if it
is nested, otherwise apply $h'$.
The resulting digests are concatenated, and then $h'$ is applied to obtain the
digest of the document (without signature).
If the container stanza contains attributes about signatures, we concatenate
the digest of the document with the public key (the value of {\em pubkey}),
the number representing the signature algorithm (the value of {\em algo}), and
the signature data (the value of {\em sig}), and apply $h'$.
The result is the digest of the signed document.

To selectively disclose the contents of a document, the parts not to be
disclosed are represented as digest elements.

\subsection{Representation in JSON}
We also propose a representation in JSON (Figure \ref{fig-json})
that can be converted to the XML data format.
The {\em salt} object is used for specifying salts for the corresponding items.
For partial disclosure, a name starting with ``digest''
such as ``digest1'' and ``digest2''
(as names within an object need to be unique
in JSON specification) can be put in place of
each item to hide with the
digest of the corresponding XML element as the value, in which case the
corresponding name-value pair in the {\em salt} object is removed.
The {\em algo}, {\em sig}, {\em pubkey} members are translated into the
respective attributes of the container XML
stanza.

The {\em proof} object specifies how the existence of the document is proven
(see section \ref{sec-poe}).
The {\em spec} object specifies the blockchain information.
The {\em subtree} object is a sequence of
$\langle$position, digest$\rangle$ pairs
for reproducing the Merkle
tree.
A position is either ``left'' or ``right'', indicating whether the digest
should be placed on the left or right side for concatenation when computing
from the digest of the (signed) document as a leaf of the Merkle tree down to
the Merkle root.

\begin{figure}[h!]
\begin{center}
{\scriptsize{\tt
\begin{lstlisting}[frame=single]
{
  "id": "4445 4558 1689",
  "name": "Wednesday Addams",
  "born": "1980-02-12",
  "address": "0001 Cemetery Lane",
  "public-key": "04d49e0786a37efce8552d6fd156...",
  "salt": {
    "id": "rGheZ.V8hqDtFw4Hy!GG",
    "name": "!f@AKAYeC63zGfMwdxtm",
    "born": "Z.uRox.GB7B3dGxPzkjB",
    "address": "h.g!-DKr.8P@EXE9QhxY"
  },
  "algo": "ecdsa-p256v1",
  "sig": "deec149de8e3055815fbca7d8719cb4fe871...",
  "pubkey": "04d49e0786a37efce8552d6fd1566d7c...",
  "proof": {
    "spec": {
      "subsystem": "ethereum",
      "network": "ropsten",
      "contract": "BBcAnchor",
      "contract_address": "0x43fA68173D1E1AFA3D...",
      "block": 9728235
    },
    "subtree": [
      {
        "position": "left",
        "digest": "9c2b8ea3291910a40eed5a62bd50..."
      },
      :
    ]
  }
}
\end{lstlisting}
}}
\caption{Sample JSON structure.}\label{fig-json}
\end{center}
\end{figure}

\subsection{Proof of Existence}\label{sec-poe}
\begin{figure*}[h!]
\begin{center}
\includegraphics[scale=0.6]{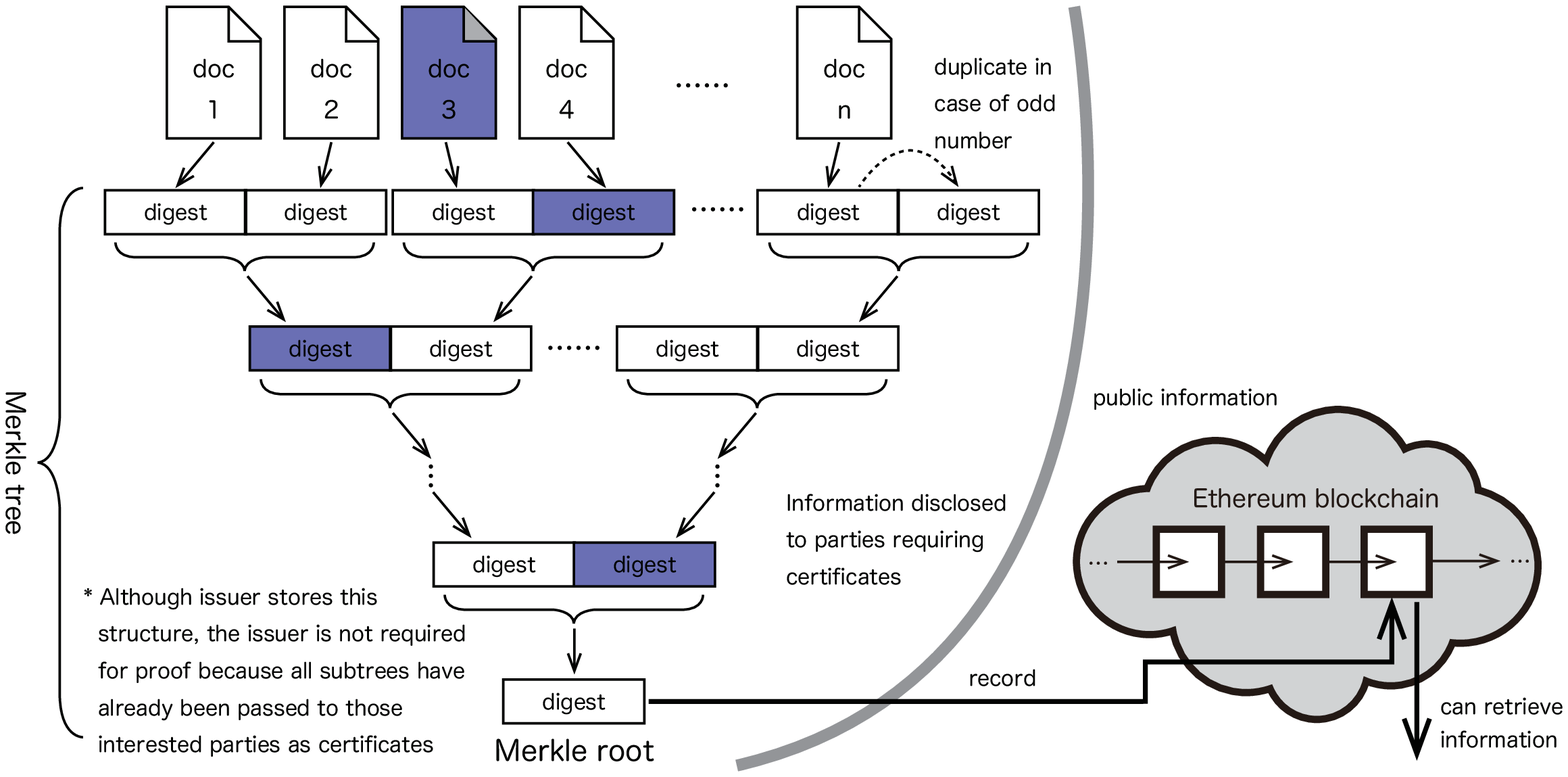}\\
{\footnotesize
\begin{itemize}
\item[*] Starting with the digest of doc 3 (calculated as shown in
Figure \ref{fig-xml}), verifier will know the series of digests to be
concatenated, so they can reproduce the calculations
down to the Markle root,
and confirm that the root matches the value recorded in the Ethereum smart
contract.
\end{itemize}
}
\caption{Merkle tree for proof of existence.}\label{fig-merkle}
\end{center}
\end{figure*}

We use a Merkle tree for proof of existence of multiple documents registered
at one time for efficiency, as illustrated in Figure \ref{fig-merkle}.
We can use a smart contract on Ethereum as described in a previous
work\cite{watanabe2020proof}
that we show in Figure \ref{fig-bbcanchor},
which stores a mapping between Merkle roots and
heights of the blocks at which the roots are stored, so that verifiers can
know the approximate timing of the registration.

\begin{figure*}[h!]
\begin{center}
{\notsotiny{\tt
\begin{lstlisting}[frame=single]
contract BBcAnchor {
    mapping (uint256 => uint) public _digests;

    constructor () public {
    }

    function getStored(uint256 digest) public view returns (uint block_no) {
        return (_digests[digest]);
    }

    function isStored(uint256 digest) public view returns (bool isStored) {
        return (_digests[digest] > 0);
    }

    function store(uint256 digest) public returns (bool isAlreadyStored) {
        bool isRes = _digests[digest] > 0;
        if (!isRes) {
            _digests[digest] = block.number;
        }
        return (isRes);
    }
}
\end{lstlisting}
}}
{\footnotesize
\begin{itemize}
\item[*] This contract saves the current block number for a stored digest.
\end{itemize}
}
\caption{BBcAnchor smart contract code (excerpt) to store and
check the existence of Merkle roots}\label{fig-bbcanchor}
\end{center}
\end{figure*}

To verify a proof, the verifier reproduces the Merkle tree from the digest of
the (signed) document and the subtree, and checks if the Merkle root is stored
in the contract on the Ethereum blockchain.

\section{Evaluation}\label{sec-evaluation}
Our proposal apparently has advantages over the known practice we described in
secion \ref{sec-known-practice} with respect to handling nested structured
documents and partial signatures.

For computational efficiency, we show a comparison of the number of digest
calculations before signing in Table \ref{tab-eval}, assuming a flat document.
If the structure is nested, the results are recursively applied.
If all items are  disclosed, the digest calculation count for verification
equals that for registration.
Upon sending out the document data for verification, the same calculation
as registration is needed unless all interim digests are pre-calculated and
stored.

To be fairer, we can approximate the cost of the final digest calculation for
our proposal before signing as $\frac{n}{2}$ instead of $1$, because $n$
digests need to be concatenated for our proposal instead of $2$ for the
interim calculation of a Merkle tree\footnote{The real cost should depend on
the hash function being used.}.
Then we obtain $\frac{3n}{2}$ for registration and $d + \frac{n}{2}$ for
verification.
Still, our proposal has an advantage for registration or equivalent
calculations where $n > 2$.
For verification, the known practice may look better with small $d$ and large
$n$, but the document would typically be nested for large $n$, in which case
our proposal would have an equal or larger logarithmic effect.

\begin{table}
\begin{center}
\caption{Comparison of the number of digest calculations.}\label{tab-eval}
{\small
\begin{tabular}{lcc}\hline
&
\multicolumn{1}{c}{Our Proposal}&
\multicolumn{1}{c}{Known Practice}\\\hline
Registration&$n + 1$&$\geq 2n - 1$\\\hline
Verification&$d + 1$&$\geq \log_2{n} + 1^*$\\\hline
\end{tabular}
}
{\footnotesize
\\$n$: number of items $\;\;\;\;$ $d$: number of disclosed items\\
* lower bound where $d = 1$
}
\end{center}
\end{table}

\section{Related Work}\label{sec-related}
\subsection{Blockchain-based Selective Disclosure}
\cite{9343074} is an example of an application of the known practice using a
Merkle tree to represent a credential, where the Merkle root is digitally
signed.

Coconut\cite{sonnino2020coconut} provides unlinkable selective disclosure,
but with higher cost.
Verification, for example, cost {\em gas},
or transaction fee,
in their Ethereum implementation,
whereas our proposal does not require {\em gas} for verification.

Blockcerts\cite{blockcerts:web}
standard, aligned with IMS Open Badges\cite{OpenBadges:web} and W3C Verifiable
Credentials\cite{Zundel:19:VCD}, is often used for implementing certificates
with blockchain, but selective disclosure is considered future work.

There are proposals, \cite{JOHARI2021} for example, to use blockchain to store
personal privacy information such as medical records in a verifiable form.
It is hoped that selective disclosure will be realized in such situations.

\subsection{W3C Recommendations}\label{sec-w3c}
Verifiable credentials data model\cite{Zundel:19:VCD}, a recommendation by W3C
and its implementation guidelines\cite{Sambra:19:VCI} suggest that unique
presentations of the credential should be generated for selective disclosure
to achieve unlinkability.
The implication of this for our proposal is to change salts to generate
provable documents on a case-by-case basis with help from the issuer.

\section{Discussion}\label{sec-discussion}
\subsection{Strengths of Our Proposal}
Our proposal has the advantage of being computationally simple compared to
the known practice and other related work.
Another advantage is that it is easy to handle structured documents, and
allows digital signatures on parts of the documents.

\subsection{Weaknesses of Our Proposal}
One problem is that a document can only be selectively disclosed at the
granular level of hiding a section that is expressed as an XML element.
This may not be a problem in processing ID cards, but it can be a problem if,
for example, we want to prove the authenticity of a requested public document
while the individuals' names need to remain hidden for defined years.

There is
also
a linkability issue with our proposal, because the digest to be
signed is always the same for the same document (if the same salts are used).
Issuers alone cannot track document holders because issuers are not required
for the verification process.
But verifiers can.
For example, a liquor shop would know that the same person came to shop for
two consecutive days if Wednesday Addams did so.

\subsection{How to Overcome the Weaknesses}\label{sec-overcome}
The problem of granularity can be solved by applying structured documents,
which is a strength of our proposal.
We can prepare a section defining a substitution string for any
sensitive word or name such that ``Wednesday Addams'' is placed as
``$\backslash$1'',
for example, and use ``$\backslash$1'' in the document instead.
The section for the substitution string can then be hidden when necessary.
This solution requires no changes in the proposed technique itself.

The
linkability
problem can be resolved either by following the recommendations by W3C
(see section \ref{sec-w3c}), or by making legal arrangements to make tracking
of document holders illegal for verifiers.

While privacy is certainly important, it is sometimes necessary to be able to
track the actions of specific individuals (criminals, victims or missing
persons) in order to maintain public order and human safety.
Especially with regard to certificates issued by public institutions, there
may be reasons to leave the possibility of tracking by investigative
authorities (this should require a court order), who are also public
institutions.
We believe that thorough privacy should be achieved through other means.

\section{Conclusions}\label{sec-conclusion}
\subsection{Summary}
We proposed a lightweight solution for selective disclosure of part of
documents for privacy protection that allows partial signatures for assuring
authenticity, where existence of documents is proven efficiently through use
of blockchain.
Although we only showed a JSON representation for front-end processing, it
should be applicable to other representations such as YAML.

Our implementation is available at
GitHub\footnote{
https://github.com/beyond-blockchain/examples/tree/master/certify-web}
as a sample web application and APIs.

\subsection{Future Work}
To avoid the problem of linkability, it may be necessary to devise a protocol
that passes the presentation of credentials to the user on a case-by-case
basis, following the recommendations by W3C that a different presentations of
credentials should be passed for different occasions to verifiers for
protecting privacy of users.

Active involvement in standardization can also be our future work.

\section*{Acknowledgment}
This work was supported by JSPS KAKENHI Grant Number JP21H04872.

\bibliographystyle{plain}
\bibliography{selective-disclosure}

\begin{thebibliography}{10}

\bibitem{blockcerts:web}
blockcerts.org.
\newblock {Blockcerts : The Open Standard for Blockchain Credentials}.
\newblock https://www.blockcerts.org.

\bibitem{10.1145/775152.775176}
Laurence Bull, Peter Stanski, and David~McG. Squire.
\newblock {Content Extraction Signatures Using XML Digital Signatures and
  Custom Transforms On-Demand}.
\newblock In {\em Proceedings of the 12th International Conference on World
  Wide Web}, page 170–177, 2003.

\bibitem{Buterin2013:Ethereum}
Vitalik Buterin.
\newblock {A Next-Generation Smart Contract and Decentralized Application
  Platform}, 2013.
\newblock $\:$\\https://github.com/ethereum/wiki/wiki/White-Paper.

\bibitem{10.1007/978-3-540-28628-8_4}
Jan Camenisch and Anna Lysyanskaya.
\newblock {Signature Schemes and Anonymous Credentials from Bilinear Maps}.
\newblock In {\em Advances in Cryptology -- CRYPTO 2004}, pages 56--72, 2004.

\bibitem{Hitchens2018:Merkle}
Rob Hitchens.
\newblock {Selective Disclosure in Solidity for Ethereum}, 2018.
\newblock
  $\:$\\https://medium.com/robhitchens/selective-disclosure-with-proof-f6a1ac7be978.

\bibitem{OpenBadges:web}
{IMS Global Learning Consortium Inc.}
\newblock {IMS Open Badges}.
\newblock https://openbadges.org.

\bibitem{JOHARI2021}
Rahul Johari, Vivek Kumar, Kalpana Gupta, and Deo~Prakash Vidyarthi.
\newblock {BLOSOM}: Blockchain technology for security of medical records.
\newblock {\em ICT Express}, 2021.
\newblock $\:$(in press).

\bibitem{Merkle1988:Tree}
Ralph~C. Merkle.
\newblock {A Digital Signature Based on a Conventional Encryption Function}.
\newblock In {\em Advances in Cryptology --- CRYPTO '87}, pages 369--378,
  Berlin, Heidelberg, 1988. Springer.

\bibitem{9343074}
R.~{Mukta}, J.~{Martens}, H.~y.~{Paik}, Q.~{Lu}, and S.~S. {Kanhere}.
\newblock {Blockchain-Based Verifiable Credential Sharing with Selective
  Disclosure}.
\newblock In {\em 2020 IEEE 19th International Conference on Trust, Security
  and Privacy in Computing and Communications (TrustCom)}, pages 959--966,
  2020.

\bibitem{Nakamoto2008:Bitcoin}
Satoshi Nakamoto.
\newblock {Bitcoin: A Peer-to-Peer Electronic Cash System}, 2008.
\newblock $\:$\\http://bitcoin.org/bitcoin.pdf.

\bibitem{Saito2016:Blockchain}
Kenji Saito and Hiroyuki Yamada.
\newblock {What's So Different about Blockchain? -- Blockchain is a
  Probabilistic State Machine}.
\newblock In {\em 2016 IEEE 36th International Conference on Distributed
  Computing Systems Workshops (ICDCSW)}, pages 168--175, 2016.

\bibitem{Sambra:19:VCI}
Andrei Sambra.
\newblock {Verifiable Credentials Implementation Guidelines 1.0}.
\newblock {W3C} note, W3C, September 2019.
\newblock https://www.w3.org/TR/2019/NOTE-vc-imp-guide-20190924/.

\bibitem{sonnino2020coconut}
Alberto Sonnino, Mustafa Al-Bassam, Shehar Bano, Sarah Meiklejohn, and George
  Danezis.
\newblock {Coconut: Threshold Issuance Selective Disclosure Credentials with
  Applications to Distributed Ledgers}, 2020.

\bibitem{watanabe2020proof}
Hiroshi Watanabe, Kenji Saito, Satoshi Miyazaki, Toshiharu Okada, Hiroyuki
  Fukuyama, Tsuneo Kato, and Katsuo Taniguchi.
\newblock {Proof of Authenticity of Logistics Information with Passive RFID
  Tags and Blockchain}, 2020.

\bibitem{ZHANG202093}
Shijie Zhang and Jong-Hyouk Lee.
\newblock Analysis of the main consensus protocols of blockchain.
\newblock {\em ICT Express}, 6(2):93--97, 2020.

\bibitem{Zundel:19:VCD}
Brent Zundel, Manu Sporny, Grant Noble, Daniel Burnett, and Dave Longley.
\newblock {Verifiable Credentials Data Model 1.0}.
\newblock {W3C} recommendation, W3C, November 2019.
\newblock https://www.w3.org/TR/2019/REC-vc-data-model-20191119/.

\end{thebibliography}

\end{document}